# Anisotropic behavior of TiB$_2$ nanoparticles reinforced 2024Al composites rolling sheet

J.M. Li[a,b], J. Liu[c], L. Wang[c], Z. Chen[a*], Q.W. Shi[b,c], C.Y. Dan[c], Y. Wu[c], S.Y. Zhong[b,c*], H.W. Wang[a]

[a] *State Key Laboratory of Metal Matrix Composites, Shanghai Jiao Tong University, Shanghai 200240, PR China*

[b] *SJTU-ParisTech Elite Institue of Technology, Shanghai Jiao Tong University, Shanghai 200240, PR China*

[c] *School of Materials Science and Engineering, Shanghai Jiao Tong University, Shanghai 200240, PR China*

*Corresponding authors: Z. Chen, Email: zhe.chen@sjtu.edu.cn;
S.Y. Zhong, Email: shengyi.zhong@sjtu.edu.cn

## Abstract

The in-plane anisotropy (IPA) of mechanical properties existed in plate products is often undesired since it limits the industrial application and increases the difficulty of material processing. It is thus one of the important factors which requires strict control during the manufacture of high-performance aluminum alloy plates. The anisotropic behavior of TiB$_2$/2024Al composite rolling sheet under different heat treatment is studied in this article. T4 heat treated specimen shows good isotropy of yield strength and its texture components are significantly weakened compared with traditional Al alloys. The evolution of microstructures for T4 specimen was studied. The submicron TiB$_2$ particles increase the driving force of recrystallization during rolling process and static recrystallization (SRX) is stimulated, thus induces texture randomization. Anisotropic behavior of yield strength occurs in specimen after T3 heat treatment. It is supposed that the difference in the activation of slip system under different loading direction gives rise to this anisotropy.

## 1. Introduction

Anisotropy of the yield strength is typically referred as the resistance to yielding as a function of orientation [1,2]. The common reference to the anisotropy of plate products can be classified as [2,3] through-thickness anisotropy (TTA) mainly for the thick plates and in-plane anisotropy (IPA) for thin plates. Traditional aluminum alloys used in the aerospace field such as Al-Li alloys and 2XXX aluminum alloys were reported [2,4,5] to be severely anisotropic in terms of mechanical properties. Generally, the anisotropy in both cold-rolled and extruded sheets [6,7] is often undesired and should be avoided during manufacture procedure. For example, the plastic anisotropy has a great impact on the deep drawability of the aluminum alloys, giving rise to the well-known earing phenomenon [8,9] and flow stress anisotropy induced by dislocation substructure which could strongly affects the formability of sheets [6]. The most common sheet production processes for aluminum alloys produce anisotropic mechanical properties as a result of the crystallographic texture development in the plane of the rolling direction [12]. Since many manufacturing steps had to be tailored to accommodate the differences in properties as a function of orientation [13], the plates with anisotropy has many limitations in industrial application meanwhile it also increases the difficulty of material processing. Therefore, many techniques have been proposed to reduce the anisotropy in order to have a better comprehensive mechanical properties. It is thus one of the important indicators which require strict control during the manufacture of high-performance aluminum alloy sheets.

Considering the importance of the anisotropic behavior to the metal sheet formability, a number of yield criteria have been proposed since decades to characterize its anisotropy and to facilitate the numerical simulation. In 1948, Hill proposed the famous Hill48 yield criterion [14] to describe the yield surface of orthogonal anisotropic material. Afterwards, Hill proposed several criteria to better estimate the flow stress in the equi-biaxial tensile [15], to include a shear stress term in the criterion [16] and to minimize the difference in the description of yield strength and Lankford coefficient [17] correspondingly. However, limitation still exists when applying these criteria [18]. On the other hand, in 1972 Hosford [19] has put forward an isotropic yield criterion based on the poly-crystallographic calculation. Afterwards, Barlat et al. [20-22] have enlarged its application field by adding a shear stress term [20], considering anisotropic situation in the plane stress state [21] and full stress state [22]. This series of criteria are applicable to alloys having strong texture components, reflecting the influence of anisotropic yield strength on the flow stress and forming limits [18]. Yet, whether they are applicable for composite materials has barely been studied.

Apart from the description of materials' anisotropic yield surface, numerous studies have been developed to study the cause of anisotropy. It is believed that microstructural anisotropy, including preferential grain orientation, grain shape, directionality of dislocation substructure, preferentially aligned precipitates, gives rise to mechanical anisotropy. Hargarter et al. concluded [1] that the preferential precipitation of θ' and Ω on {100} and {111} planes has certain effects on the anisotropy of Al-Cu and Al-Cu-Mg-Ag alloys; Jata et al. experimentally found [2] that the IPA of 7XXX and Al-Li base alloys is strongly related to Brass texture intensity; Lopes et al. studied [23] influence of texture components and grain shape on the strain hardening anisotropy of 1050 Al alloy by both experiments and polycrystal simulation; researchers from Riso National Laboratory [10,24-26] revealed that the directionality of dislocation substructure, especially geometrically necessary boundaries (GNBs), has great influence on the anisotropy of different metals through a systematic study. Therefore, it is of great significance to study whether microstructural anisotropy exists in the

TiB$_2$/Al composite plates and its influence on the anisotropy of mechanical properties.

Compared with ex-situ TiB$_2$/Al composite, in-situ TiB$_2$/Al composite has a cleaner interface between TiB$_2$ particles and Al matrix [27,28]. It is also easier to fabricate and has better comprehensive mechanical properties, including a superior strength and ductility [29], a higher fracture toughness [30], an improved fatigue limit [31] etc. Considering the distribution state of particles is an additional factor which would cause microstructural anisotropy, its influence on the anisotropy of mechanical property worth discussing. Although it has been mentioned in the previous study [11], study systematical has not been performed yet. In this article, the anisotropy of yield strength of 4.2 wt.%TiB$_2$/2024Al composite sheet with different thermomechanical treatment will be assessed using the yield functions. Since its thickness is 2mm which could be approximated as plane stress state, Barlat-Lian89 criterion was selected to describe its yield surface and Hill48 criterion was chosen as a comparison result. Then the evolution of microstructure and texture will be characterized with neutron diffraction technology and Electron Backscattered Diffraction (EBSD) technology. The microstructural cause on the anisotropic behavior were discussed.

## 2. Experimental procedure

The chemical composition of the 4.2 wt. %TiB$_2$/2024Al composite plate used in this study is shown in Table 1. The size of TiB$_2$ particle ranges from 20 to 500 nm, measured by the restricted moment method [32]. The composite ingots were fabricated in-situ reaction methods and then followed direct chilled casting with the detail of fabrication could be found in previous research [33]. The cast raw materials were then rolled in West Cinda Aluminum industry Co. Ltd with its final thickness of 2mm. The solution treatment was performed at 505 ℃ for 50 minutes followed by water quenching. Some specimens were immediately pre-stretched along the rolling direction to a strain of 2 %, 4 % and 8 %, correspondingly. Then these specimen are subjected to natural aging for more than 96 h. After that the tensile specimens were obtained by electro-discharge machining according to ASTM E8M-13a, 2013 [34](see figure 1a) at different angles θ to the rolling direction with 15 ° interval (i.e. θ=0°, 15 °, 30 °, 45 °, 60 °, 75 ° and 90 °) in order to study its anisotropic behavior. The sampling method is showed in figure 1b.

Uniaxial tensile tests were conducted at room temperature with strain rate $10^{-4}$ /s to obtain the typical stress-strain curves on a Zwick/Roell Z020 test machine. Microstructure analysis was performed using EBSD technology on NOVA nano SEM 230 with an accelerating voltage of 20 kV and the step size 0.9 μm. Before EBSD characterization, all specimens were mechanically polished. An additional ion milling was conducted by Leica EM TIC 3X. The corresponding figures and data were proceeded with the HKL channel 5 software. In the corresponding calculation, misorientations larger than 15 ° were defined as high angle grain boundaries (HAGBs), shown in blacklines while those in the range of 2 ° and 15 ° defined as low angle grain boundaries (LAGBs), shown with grey lines. Schmid factor of each grain under different loading directions was calculated and mapped with the help of HKL channel 5 software. Texture measurements were carried out at laboratory CMRR, Institute of Nuclear Physics and Chemistry, Mianyang, China using neutron diffractometer Kylin with a wavelength of 0.231 nm. Specimens cut from the middle sections of rolling sheets were stacked to form 1x1x1 cm³ cube volume. Three individual (1 1 1), (2 0 0) and (2 2 0) pole figures were measured for each sample which correspond to 5 ° step along the azimuthal and polar angles.

## 3. Results

*3.1 Anisotropic behavior of yield strength*

Typical stress-strain curves of 4 differently heat-treated specimens oriented at 0 ° (longitudinal direction, noted LD in the figure) , 45 ° (diagonal direction, noted DD in the figure) and 90 ° (transverse direction, noted TD in the figure) from rolling direction obtained from uniaxial tensile test were shown in the figure 2. Curves of the T4-treated specimen for 3 directions almost superpose, indicating the excellent isotropy of T4 treated specimens. As for T3-treated specimens, anisotropy is observed. The specimen of rolling direction has the highest strength and lowest elongation while the 45 ° specimen is on the contrary. Yield strengths of differently heat-treated specimens in the same direction increase with the pre-stretching magnitude meanwhile the elongation decrease. No apparent necking phenomenon is observed in any of the specimen. Compared with T4 specimen, ultimate tensile strength also increase with the pre-stretching magnitude (25 MPa increment in the rolling direction after a 8 % pre-stretching magnitude), but to a less extent compared with the yield strength (176 MPa in the rolling direction after a 8 % pre-stretching magnitude). Furthermore, serrated stress-strain curve is observed for T3 specimen. It is due to the interaction between mobile dislocation introduced during pre-stretching procedure and solute atoms [35]. This phenomenom is not in the scope of this article.

In order to further study the anisotropic behavior of yield strength and determine the yield functions' parameters, different specimens along 7 directions were tested with their results showed in the figure 3. The blue, green, black and red lines stood for the T4 specimens, T3-2% specimens, T3-4 % specimens and T3-8 % specimens correspondingly. These results of yield strengths were in good agreement with the above results, indicated that T4 heat-treated specimens demonstrated pretty well isotropy of yield strength while more anisotropy was introduced with increasing pre-stretching magnitude.

*3.2 Microstructure and texture*

The microstructures of section of longitude direction and transverse direction (noted as LD-TD section in the following part) for T4 treated specimen and T3-8 % treated specimen were showed in figure 4 and figure 5. BSE images revealed that TiB$_2$ particles were uniformly distributed in both T4 and T3-8 % specimens with no apparent directionality. Distribution state of TiB$_2$ particles in the T4 and T3-8% samples showed no obvious difference. Particles displayed faceted shape and their size were various. No significant particle aggregation was observed in either samples. Systematical study of size distribution of TiB$_2$ particles was performed in previous research [32], which suggested that the size of most TiB$_2$ particles are in the range of 20-500 nm, with some of particles as large as 1μm. Figure 5 gives inverse pole figures (IPFs) of LD-TD section from EBSD analysis for T4 and

T3-8% specimen, with its legend showed in the right top corner. The X direction is parallel to the rolling direction. Black lines represent high angle grain boundaries (HAGBs, θ>15°) and yellow lines represent low angle grain boundaries (LAGBs, 15°>θ>2°). Grains of both specimen are equiaxed with their aspect ratio of the same level. No obvious preferential orientation was shown in either sample. Large amount of LAGBs could be observed in T3 speciemen (figure 5b) compared with almost none LAGBs in T4 specimen (figure 5a). This is due to the dislocations introduced during pre-stretching process knowing that the dislocations could rearrange and form LAGBs.

From figure 5-b directional arrangement of LAGBs could be observed in part of grains of the T3 specimen. Therefore, grains with and without the directional arrangement of LAGBs were randomly picked out to have a better analysis. Figure 6 shows the grains whose orientations are close to <001> and <111> respectively a) with obvious grain rotation trace and b) without them. Each grain was numbered to facilitate further discussion in the following part. The Schmid factor of each grain loaded at LD, DD and TD direction were calculated using HKL channel 5 software and indicated below in the figure. The directions with largest Schmid Factor for each grain are highlighted in the figure. The calculation results revealed that loading in LD direction would have the smallest Schmid Factor for all grains while in almost all grains loading in DD would lead to the largest Schmid Factor. More specifically, the difference between Schmid Factor under different loading direction of grains close to <111> orientation (blue grains) is larger than those close to <001> orientation (red grains), especially for those small-size grains. Considering the grains close to <111> orientation accout for a large portion (see figure 5-b) and most of them are small grains, this might have an influence on the macroscopic anisotropy of its yield strength.

Bulk texture measurement results of T4 and T3-8% specimens were shown in the form of pole figures (PFs, see figure 7). No preferential grain orientation was shown in both specimens. The maximum multiple uniform density (MUD) of 4.98 and 3.06 for T4 and T3-8% sample indicating that neither of the specimen shows the typical texture components that could be observed in traditional Al alloys.

*3.3 Description of the yield surface*

Yield surface usually takes the form of an implicit equation [36]:

$$\varphi(\overline{\sigma}, Y) = 0 \tag{1}$$

where $\varphi(\overline{\sigma},Y)=\overline{\sigma}-Y$, $\overline{\sigma}$ is the equivalent stress and $Y$ is, in practical, considered as the uniaxial tensile yield strength along the rolling direction [37]. The equivalent stress $\overline{\sigma}$ is described with the help of yield criteria, two of which were chosen in this paper: Hill 1948 yield criterion and Barlat 1989 yield criterion.

--Hill's 1948 yield criterion could be described with four parameters a, b, c and d:

$$\overline{\sigma} = \sqrt{a\sigma_{xx}^2 + b\sigma_{yy}^2 + c(\sigma_{xx} - \sigma_{yy})^2 + 2d\sigma_{xy}^2} \tag{2}$$

This criterion is a more general form of von Mises criterion for orthogonal anisotropic materials and thus was widely used due to its simplicity and its considerable ability to describe the behavior of orthogonal anisotropic sheet metals.

-- Barlat's 1989 yield function takes the following form with also four parameters a, m, h and p:

$$\bar{\sigma} = \left(\frac{a|K_1 + K_2|^m + a|K_1 - K_2|^m + (2-a)|2K_2|^m}{2}\right)^{1/m} \quad (3)$$

where:

$$K_1 = \frac{\sigma_{xx} + h\sigma_{yy}}{2} \quad (4)$$

$$K_2 = \sqrt{\left(\frac{\sigma_{xx} - h\sigma_{yy}}{2}\right)^2 + p^2\sigma_{xy}^2} \quad (5)$$

This criterion could describe the sheets' anisotropic behavior of yield strength with strong texture and reflect comprehensively the effect of IPA to the formability of sheets. According to previous research [37-39], the value of $m$ is recommended to be 8 for FCC metals.

The parameters of yield criterion were determined from experimental results of yield strength from the uniaxial tensile test along four different directions: in the Hill-1948 yield criterion and Barlat-1989 yield criterion, $\sigma_{xx}$, $\sigma_{yy}$ and $\sigma_{xy}$ could be obtained from the experimental result:

$$\begin{bmatrix} \sigma_{xx} & \sigma_{xy} \\ \sigma_{yx} & \sigma_{yy} \end{bmatrix} = Q^T \cdot \begin{bmatrix} \sigma_\theta & 0 \\ 0 & 0 \end{bmatrix} \cdot Q \quad (6)$$

where $Q = \begin{bmatrix} \cos\theta & -\sin\theta \\ \sin\theta & \cos\theta \end{bmatrix}$, θ signifies the intersection angle between the loading direction and the rolling direction and $\sigma_\theta$ represents the yield strength obtained from uniaxial tensile test along loading direction. In consequence, four parameters could be solved with four equations listed with experimental data.

In addition, the equivalent elongation can be simulated using the same criterion with different parameters. Therefore, the yield surface of four different types of specimens with the information of its elongation could be plotted with the help of Matlab (showed in the following figures). In the figure 8, horizontal ordinate $\sigma_1$ stands for the stress component in the rolling direction of yield stress while vertical ordinate $\sigma_2$ stands for stress component in the transverse direction. The curves with different colors represent simulation results of yield surface of different specimens while red points represent the experiment results. Different colors represent the simulated elongation variation along all directions with its scale on the right. The difference between maximum and minimum of elongation scale were all taken for 7 % and the isotropic yield functions for 4 types of specimens was drawn with the black lines for the purpose of visualizing the evolution of its anisotropy for elongation and yield strength respectively. Both criteria reveal the evolution of yield strength to a certain extent: the nearly circular yield surface of T4 specimen indicates its isotropy while the shape becomes more elongated due to the anisotropy introduced by the increase of pre-stretching magnitude.

# 4. Discussion

*4.1 Comparison of yield criteria*

In between two yield criteria chosen in this article, Barlat89 describes more accurately anisotropic behavior of yield strength for metal sheets with strong texture components [18,21] while Hill48 describes simple orthogonal anisotropic materials. Therefore, the accuracy of two criteria in description of yield surface reveals its texture components intensity to some extent. In order to compare the accuracy of two criteria between experimental results and simulation data, root mean square error between simulation results and experimental results (noted as ε) was introduced here and calculated who takes the form of:

$$\varepsilon = \sqrt{\frac{\sum_{i=1}^{7}(\sigma_{e,i} - \sigma_{s,i})^2}{7}} \tag{7}$$

where $\sigma_{e,i}$ are experimental results and $\sigma_{s,i}$ are simulation results. Calculation result of $\varepsilon$ for two yield criteria is given in the table 3. According to the result of neutron diffraction, no obvious texture components is observed in T4 sample, so it is reasonable that Barlat89's prediction is less accurate than Hill48 for T4. As for T3-2% sample, the experimental results of 7 direction fluctuate significantly. Since Hill48 give a circular or elliptical yield surface and Barlat89 could describe the fluctuation in flow stress, this might be the reason why $\varepsilon$(Hill48) is much bigger than $\varepsilon$(Barlat89). It is worth noticing that the accuracy of Barlat89 and Hill48 in prediction of yield surface is of the same level for T3-4% and T3-8%. In Barlat89 yield function, shear stress term is included and Barlat designed this criterion to describe textured polycrystalline sheet metal [21]. As for traditional aluminum with strong texture, Barlat89 should be much more accurate in describing its yield surface compared to Hill48 [19,37], yet this is not the case in the present work. This result indicates that texture components are not as strong as traditional Al alloys reported in the literature even for T3 samples, which also verifies the neutron diffraction result of T3-8% sample.

*4.2 Recrystallization and texture randomization*

Generally speaking, recrystallization could be divided into static recrystallization (SRX) and dynamic recrystallization (DRX). SRX occurs when deformed metals are annealed above recrystallization temperature while DRX occurs during the severe plastic deformation or high deformation temperature. The grains of T4 specimen have the features of recrystallized grains, such as its dislocation free characteristic. Hence the relation between uniformly distributed TiB$_2$ particles and recrystallization is worth discussing. Numerous studies have been carried out to investigate the effects of second phase particles on the recrystallization [40-45]. It is well accepted that coarse non-deformable particles would induce recrystallization [40,41] while fine particles could either accelerate or retard recrystallization depending on deformation degree, interparticle spacing and annealing time [41,46,47]. Since rolled state composite has gone severe plastic deformation, the resolution rate on the sample surface was not high to apply EBSD technology. Therefore, here the microstructure of rolled state composite is showed using polarized optical microscopy (POM) after anode coating (see figure 9). The grains before solution treatment have pancake-like shape with its longer side parallel to the rolling direction. This result indicates that sample grains at rolled state were not largely recrystallized, i.e. the DRX is not dominant in the whole recrystallization process. However, the TiB$_2$ particles, whose size range from 25 nm-500nm according to our previous research [32], are able to act as dislocation sources. Since it was uniformly distributed in this rolling plate (figure 5-a,b), enormously high density of dislocation sources were contained in this material. It was suggested by Chopra that [45] for particle reinforced materials, formulation of nuclei should be much easier after severe initial deformation because of regions having large misorientations. In the present work, the pancake-like grains after rolling suggested that uniformly distributed fine TiB$_2$ particles has not sitmulated DRX during rolling process. Instead, they introduced large amount of dislocations by Orowan mechanism during severe rolling reduction. The driving force for recrystallization has thus increased greatly. After the next step, that is, the solution treatment process at 505 ℃, SRX occurred to release great driving force. The pancake-like dofromed grains transformed to recrystallized grains in this process. Although this high solution treatment temperature induced SRX, it also coarsened the grains to some extent due to the high annealing temperature. That is the reason why the grain size is larger compared with our previous research [11,29,30,48]. The recrystallization situation of LD-TD plane was showed in figure 10, red, yellow and blue grains stand for the deformed, recovered and recrystallized grains correspondingly. Up to 90.2% amount of grains were recrystallized grains according to the EBSD calculation.

The recrystallized grains are not randomly oriented: in a monocrystal, its grain orientation is generally rotated by a large angle around the transverse direction or normal direction on the basis of the orientation of the deformed matrix [49] and in a polycrystal, the recrystallized grains share the same forming mechanism with one of monocrystal [50]. However, since each nucleus has a certain misorientation with matrix grains and the misorientation between different grains is pretty large originally, in this way, it is difficult to show a preferential orientation for all the grains, thus textures are randomly distributed through macroscopic view. As a result, volume fraction of rolling textures in T4 specimen in the present work (less than 5%) is much lower than those results reported in the documents, considering rolling texture such as S and Brass usually accounts for [51,52] 30 %-40 % volume fraction in the typical 2XXX alloys. As for T3 specimen, they were pre-stretched by a small strain (2%-8%) after annealing at 505 ℃. Such small deformation should have little influence in the variation of texture components, so it is reasonable to see that no obvious texture was detected in T3 specimen.

*4.3 Schmid factor analysis*

As having discussed in 4.2, there should be no significant difference in texture components for T4 and T3 specimen. Basically, the anisotropy of yield strength is the result of the anisotropy in the difficulty of the activation of slip systems loaded in different sample direction. In order to analyze the anisotropic behavior of T3 specimen, Schmid factor analysis was performed as it is a routine method to analyze the anisotropy caused by slip situation, knowing that the larger the Schmid factor is, the easier the slip system is to start. According to Schmid's law, the uniaxial tensile yield stress of a monocrystalline could be expressed as:

$$\sigma = \frac{\tau_c}{m_{max}} \tag{8}$$

where $m = cos\varphi \, cos\lambda$ is called Schmid factor. Here $\tau_c$ represents the critical resolved shear stress of the monocrystalline and $\varphi$, $\lambda$ stand for intersection angle between loading direction and slip direction, loading direction and normal direction of slip plane respectively.

The microstructures of LD-ND section for T4 and T3-8% specimen were additionally investigated in order to perform Schmid factor analysis since IPA is produced as a result of the crystallographic texture development in the plane of the rolling direction [12]. With the help of HKL channel 5 software, the Schmid factor of each grain's most preferential slip system along different loading directions could be calculated and mapped. Part of calculation results for both specimens were given both in the form of color scale (figure 11) and statistics (in the middle) as an example.

The area-weighted mean of Schmid factor's reciprocal was calculated to verify its anisotropy of yield strength (see table 4) and compared with the experimental results of uniaxial yield strength in figure 12. From figure 12 we could see that the evolution of Schmid factor and experimental result of yield strength shared the same tendency for both specimen, only small differences existed between them. This phenomenon reveals that when loading at different directions, the difficulty of the activation of slip systems are almost the same for T4 specimen while those of T3 specimen do varies. This microscopic orientation-dependent property would result in the anisotropy of yield strength. It is thus of great interest to further investigate the reason why the activation of slip systems is orientation dependant for T3 specimen.

The discussion above analyzed the general slip activation situation of the sample, i.e. several hundreds of grains in total. It is also interesting to investigate the activation of slip system in single grains. According to calculated result of the Schmid Factor, most of the grains shared the same tendance that yield strength in LD direction is the highest while that in DD direction is the lowest. To further study the slip situation in different types of grains, grain 1, 3, 5 and 9 were selected here considering their different grain orientation and grain size. The Schmid factor of all slip systems in four grains mentioned above were calculated using the definition of Schmid Factor: $m = cos\varphi\, cos\lambda$ knowing the slip system of Al is {111}<110>. The calculation results is given in the table 5, with the slip direction with highest Schmid factor highlighted: the highest Schmid factor for four grains with different loading direction corresponds well to those calculated by HKL channel 5 (see figure 6). In addition, through this calculation the activated slip system could be determined directly. When loading is applied at different direction, the activated slip system for each grain is different, thus results in the difference of capacity in resisting the slip, which in a macroscopic view, is the difference of yield strength.

More specifically, the cause of anisotropy in yield strength might be different for grains close to <001> orientations (red grains in figure 7) and those close to <111> orientations (blue grains in figure 7):

1) As for red grains, several slip direction might start simultaneously. For example, when grains 5 is loaded at LD, the Schmid Factors of {-1 1 1}[0 -1 1] (0.4324), {1 -1 1}[1 1 0] (0.4287) and {1 -1 1}[0 1 1] (0.4277) slip directions have no significant differences. Similar case could be observe in grain 1 (the Schmid factor of {-1 1 1}[0 -1 1] and {1 -1 1}[0 1 1] are 0.4707 and 0.4646 respectively). Since these slip directions are in different slip planes, two or more sets of crossed slip lines would be formed. Due to the interpenetration of multiple slip systems, slip might be more difficult under this condition. On the contrary, when loading is applied in DD and TD, only one slip system could be activated. This makes them the "softer" directions compared with LD.

2) As for blue grains, the differences between largest Schmid factor and the Schmid factor of other slip system is large in all loading direction, yet the trace in the grain might have an influence on its slip behavior. Figure 13 showed the direction of activated slip plane in grain 3 when a) loaded at LD and b) loaded at DD (in order to better see its slip trace, LAGBs in these grains were removed). When loaded at LD, the activated slip system is on the plane {1 1 -1}, which forms a certain angle with the trace in grain 3 (figure 13a); when loaded at DD, the activated slip system on the plane {-1 1 1} forms a smaller angle with the trace (figure 13b). These traces was formed during pre-stretching procedure with LAGBs accumulated along them (see figure 5b). This phenomena might result in the anisotropic slip activation situation since the larger intersection angle is, the more LAGBs woud be met with the same slip length. However, further researches are needed to give out a more persuasive evidence.

## 5 Conclusion

In this article, yield strength's anisotropic behavior of in-situ 4.2 wt.%TiB$_2$/2024Al composite sheet under different heat treatment were studied. Macroscopic and microscopic characterizations were made to identify its cause. Several conclusions could be drawn as follows:

(1) Specimen under T4 heat treatment exhibited excellent isotropy for yield strength due to randomized texture component compared with traditional Al alloys. No obvious preferential grain orientation was observed.

(2) Uniformly distributed TiB$_2$ particles increased the driving force for recrystallization by introducing large amount of dislocations. SRX that occurs after the solution treatment process dominates the recrystallization process, resulting in the texture randomization.

(3) Difference in the activation of slip system in each grains of T3 treated composites cause the anisotropy despite the randomized texture components. Two types of slip activation could be concluded as follows:

    (i) For grains close to <001> direction, several slip systems were activated under LD loading while only single slip system would be activated under other loading direction, which makes LD a "harder" direction.

    (ii) For grains close to <111> direction, the difference in the intersection angle between activated slip planes and LAGBs leads to the difference in slip activation. A larger intersection angle under LD loading results in the smaller Schmid factor in this loading direction.

## Acknowledgment

This work is financially supported by the Natural Science Foundation of China [Nos. 51971137, 51701120 ,51301108, 11875192, U1930101] the Conseil Regional du Nord-Pas de Calais and the European Regional Development Fund (ERDF).

## Declaration of interest

No potential conflict of interest was reported by the authors

# Figure and table captions

Figure 1. a) Tensile test specimen according to ASTM E8M-13a; b) Daigram of sampling method

Figure 2. Stress-strain curve of a) T4 specimen; b) T3-2 % specimen; c) T3-4 % specimen; d) T3-8 % specimen in three directions

Figure 3. Anisotropic behavior of four specimens

Figure 4. SEM iamges showing $TiB_2$ particles uniformly distributed in a) T4 and b) T3 samples. Various size of $TiB_2$ particles with faceted shape were observed in c) d) both samples.

Figure 5 IPF maps of a) T4 specimen and b) T3 specimen from EBSD analysis

Figure 6 grains whose orientations are close to <001> and <111> respectively a) with and b) without obvious trace

Figure 7. Pole figure of 6wt.%$TiB_2$/2024Al sheet: a) T4 specimen ; b) T3-8% specimen

Figure 8. Yield surfaces of 4 differently heat-treated specimens. Black lines represent isotropic yield surface, colored lines represent simulation results of yield surface. Their colors represent the equivalent elongation with its scale on the right. Red points represent experimental results. Eight figures stand for yield surfaces of four kinds of specimens simulated by a)-d) Hill 48 criterion and e)-h) Barlat 89 criterion respectively.

Figure 9 POM figure of rolled state composite plate after anode coating

Figure 10 Recrystallization proportion for T4 specimen. Red, yellow and blue grains stand for the reformed, recovered and recrystallized grains correspondingly.

Figure 11. Schmid factors for a) b) c) T3-8% specimen and for d) e) f) T4 specimen under loading direction of a) d) LD; b) e) DD; c) f) TD. Calculation results of each grain shown in the middle with its legend

Figure 12 Comparison between yield strength and Schmid factor for a) T4 specimen and b) T3 specimen

Figure 13 The direction of activated slip plane in grain 3 when a) loaded at LD and b) loaded at DD. c) Illustration of change of intersection angle between slip planes and LAGBs on these trace.

Table 1. Chemical composition of 4.2 wt. %$TiB_2$/2024Al sheet

| Element | Ti | Cu | Mg | Mn | B | Al |
|---|---|---|---|---|---|---|
| Mass fraction(wt.%) | 3.15 | 4.20 | 1.80 | 0.45 | 0.95 | Bal |

Table 2 Average grain size and aspect ratio of 2 specimens

| Sample | Average grain size(μm) | Aspect ratio |
|---|---|---|
| T4 | 25.73 | 2.21 |
| T3-8% | 16.72 | 2.07 |

Table 3. The error between experimental results and the simulation of 2 yield criteria

| $\varepsilon$ | T4 | T3-2% | T3-4% | T3-8% |
|---|---|---|---|---|
| Hill48 | 4.6020 | 9.1648 | 4.9477 | 6.7017 |
| Barlat89 | 6.9101 | 6.5027 | 4.5723 | 6.9308 |

Table 4 Area-weighted mean of Schmid factor's reciprocal for different specimens

| 1/m | 0° | 15° | 30° | 45° | 60° | 75° | 90° |
|---|---|---|---|---|---|---|---|
| T4 | 2.1978 | 2.2060 | 2.1944 | 2.1839 | 2.1839 | 2.2060 | 2.2099 |
| T3-8% | 2.3026 | 2.2124 | 2.1561 | 2.1538 | 2.1580 | 2.1901 | 2.2193 |

Table 5 Schmid Factor of all slip systems in four grains

| slip plane | slip | grain 1 loaded at | grain 3 loaded at |
|---|---|---|---|

| slip plane | slip direction | LD | DD | TD | LD | DD | TD |
|---|---|---|---|---|---|---|---|
| {1 -1 1} | [1 1 0] | 0.3919 | 0.1398 | 0.3197 | 0.2426 | 0.3541 | 0.2056 |
|  | [0 1 1] | 0.4646 | 0.3381 | 0.0174 | 0.0734 | 0.0510 | 0.0020 |
|  | [-1 0 1] | 0.0727 | **0.4779** | 0.3023 | 0.1692 | 0.4052 | 0.2076 |
| {-1 1 1} | [1 1 0] | 0.3661 | 0.1514 | 0.4317 | 0.0673 | 0.2377 | **0.4557** |
|  | [0 -1 1] | **0.4707** | 0.3008 | 0.1118 | 0.2496 | 0.2329 | 0.1957 |
|  | [1 0 1] | 0.1046 | 0.4522 | 0.3199 | 0.1823 | **0.4706** | 0.2601 |
| {1 1 -1} | [1 -1 0] | 0.3980 | 0.1771 | 0.4141 | **0.4188** | 0.1723 | 0.4033 |
|  | [0 1 1] | 0.3207 | 0.0793 | 0.0328 | 0.0421 | 0.0117 | 0.0068 |
|  | [1 0 1] | 0.0773 | 0.0978 | 0.4469 | 0.3767 | 0.1606 | 0.3964 |
| {1 1 1} | [1 -1 0] | 0.3600 | 0.1141 | 0.3372 | 0.1089 | 0.4196 | 0.2580 |
|  | [0 1 -1] | 0.3146 | 0.0419 | 0.1272 | 0.1342 | 0.1936 | 0.1909 |
|  | [1 0 -1] | 0.0454 | 0.0722 | **0.4644** | 0.0252 | 0.2260 | 0.4489 |
| slip plane | slip direction | grain 5 loaded at | | | grain 9 loaded at | | |
|  |  | LD | DD | TD | LD | DD | TD |
| {1 -1 1} | [1 1 0] | 0.4277 | 0.0203 | 0.2292 | 0.2905 | 0.4682 | 0.0189 |
|  | [0 1 1] | 0.4287 | 0.4222 | 0.0669 | 0.0092 | 0.3031 | **0.4320** |
|  | [-1 0 1] | 0.0010 | **0.4425** | 0.2961 | **0.2997** | 0.1651 | 0.4131 |
| {-1 1 1} | [1 1 0] | 0.3776 | 0.0266 | **0.4951** | 0.2672 | 0.2096 | 0.0206 |
|  | [0 -1 1] | **0.4324** | 0.3131 | 0.1875 | 0.2817 | 0.3345 | 0.3977 |
|  | [1 0 1] | 0.0548 | 0.3397 | 0.3076 | 0.0145 | 0.1249 | 0.3771 |
| {1 1 -1} | [1 -1 0] | 0.4314 | 0.1294 | 0.4836 | 0.0180 | **0.4996** | 0.0154 |
|  | [0 1 1] | 0.3768 | 0.0802 | 0.1546 | 0.0252 | 0.2602 | 0.0086 |
|  | [1 0 1] | 0.0545 | 0.0492 | 0.3290 | 0.0431 | 0.2394 | 0.0068 |
| {1 1 1} | [1 -1 0] | 0.3739 | 0.0826 | 0.2407 | 0.0053 | 0.1782 | 0.0549 |
|  | [0 1 -1] | 0.3732 | 0.0289 | 0.0998 | 0.2473 | 0.2287 | 0.0257 |
|  | [1 0 -1] | 0.0008 | 0.0536 | 0.3405 | 0.2420 | 0.0506 | 0.0291 |

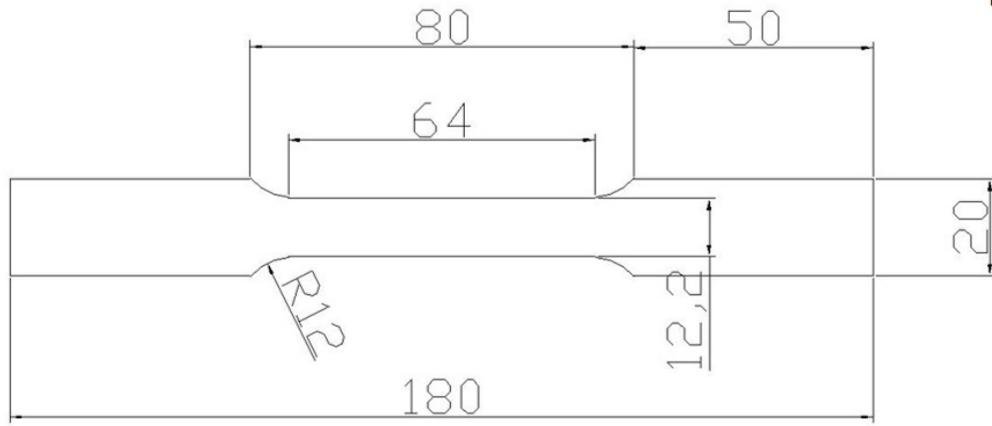 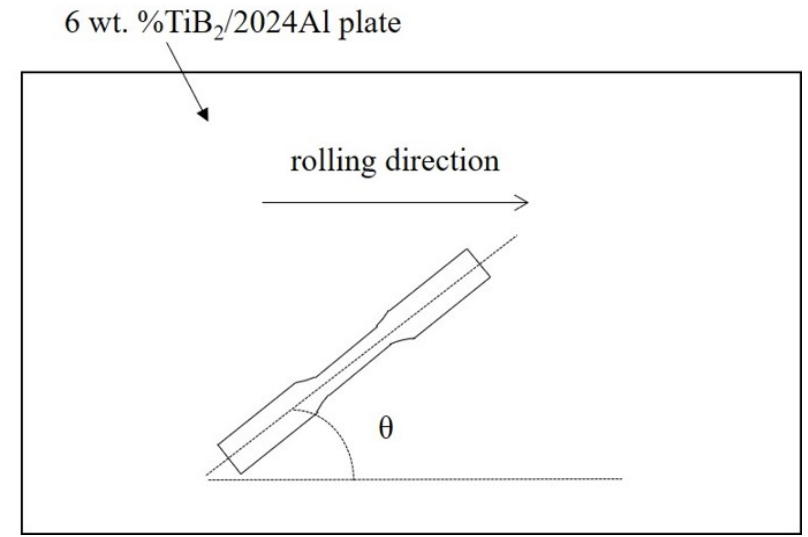

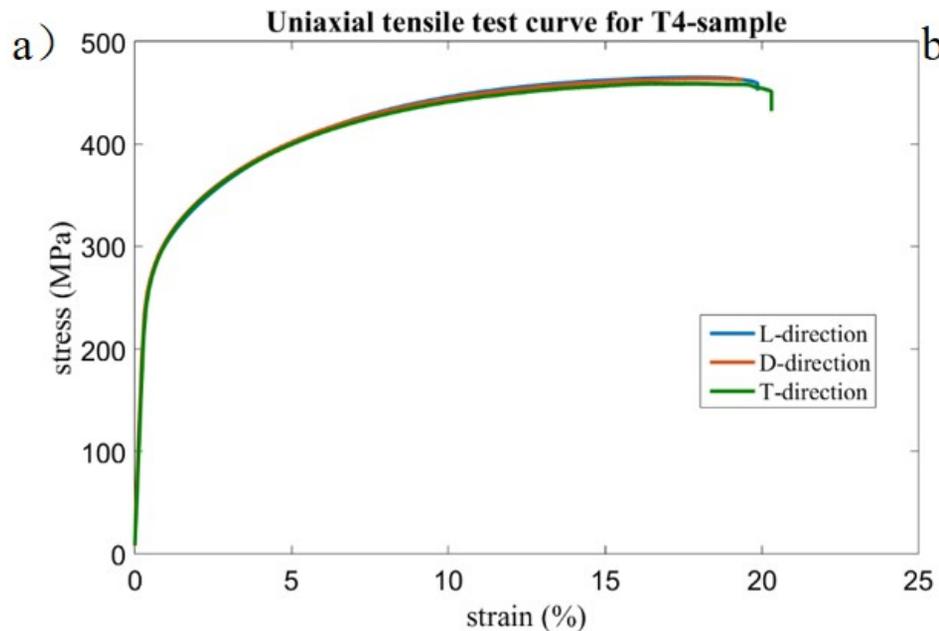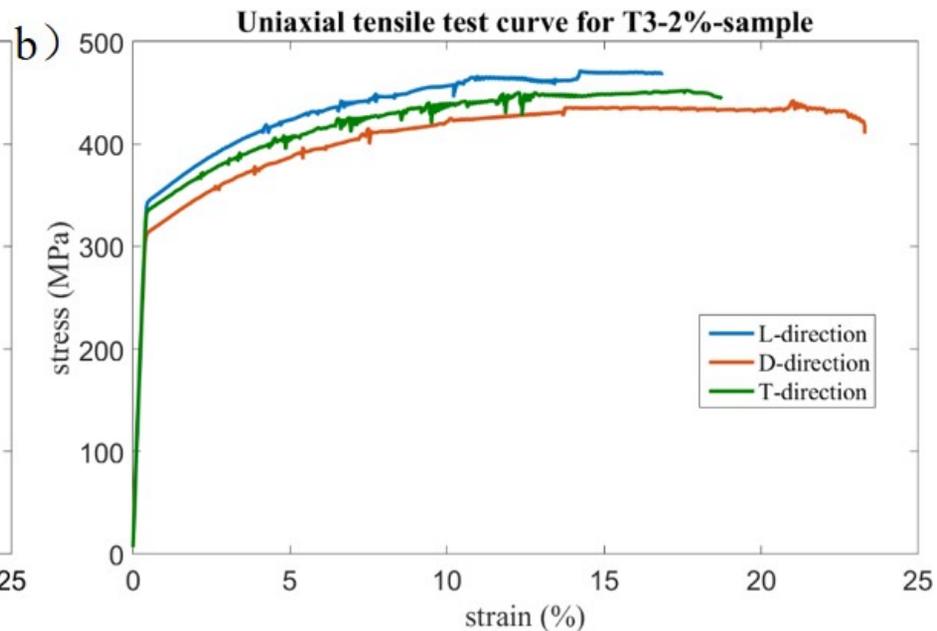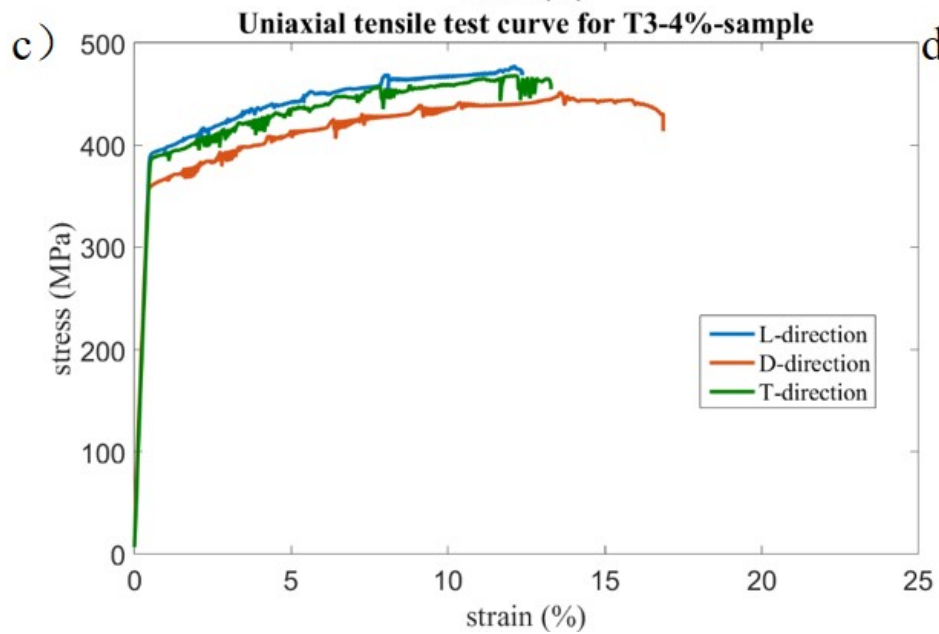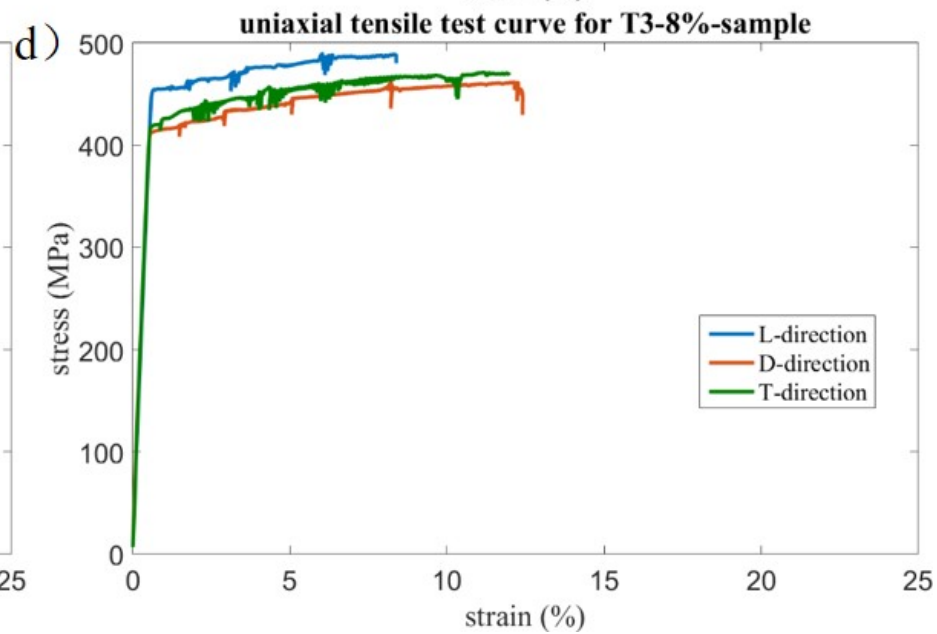

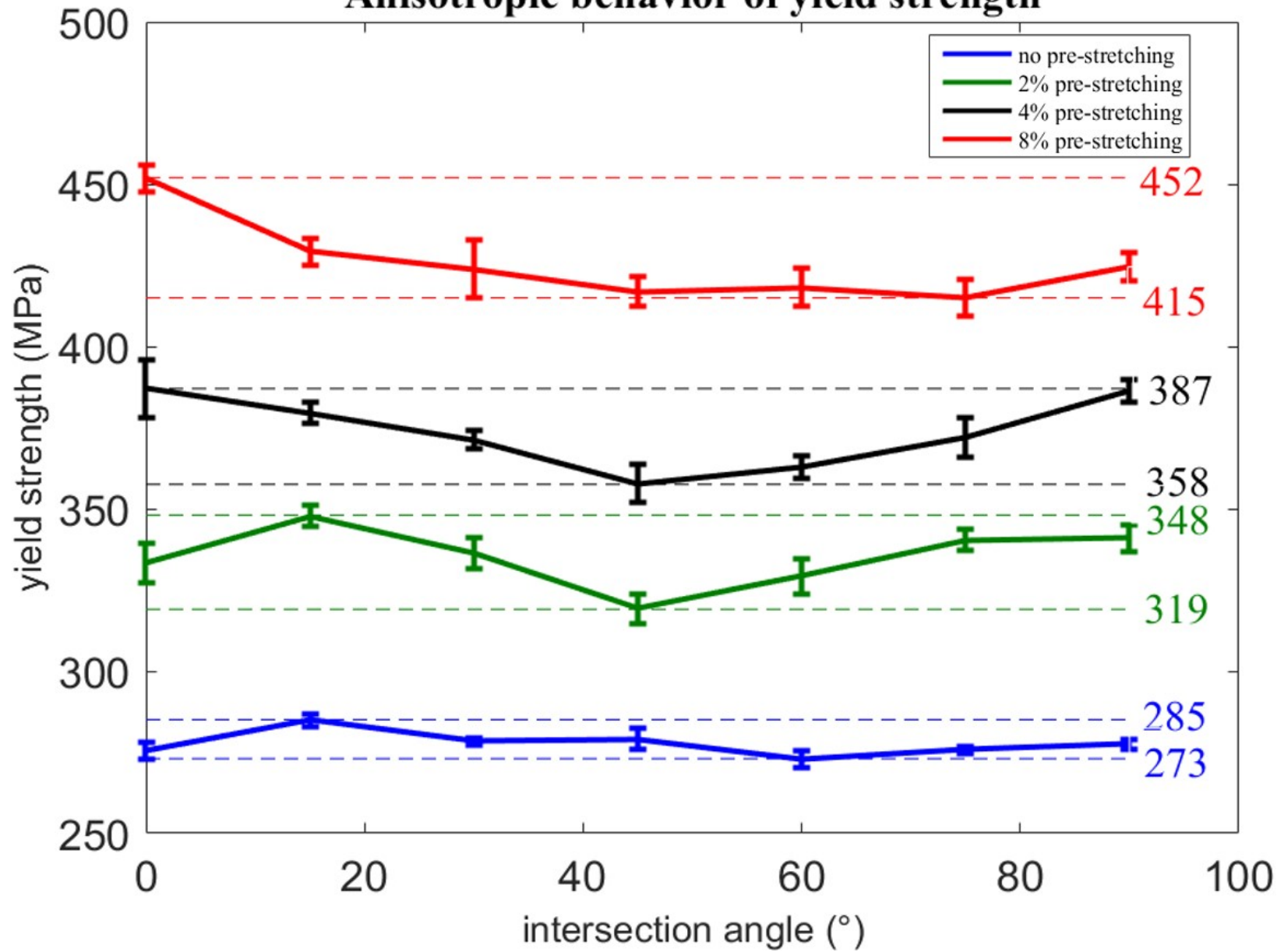

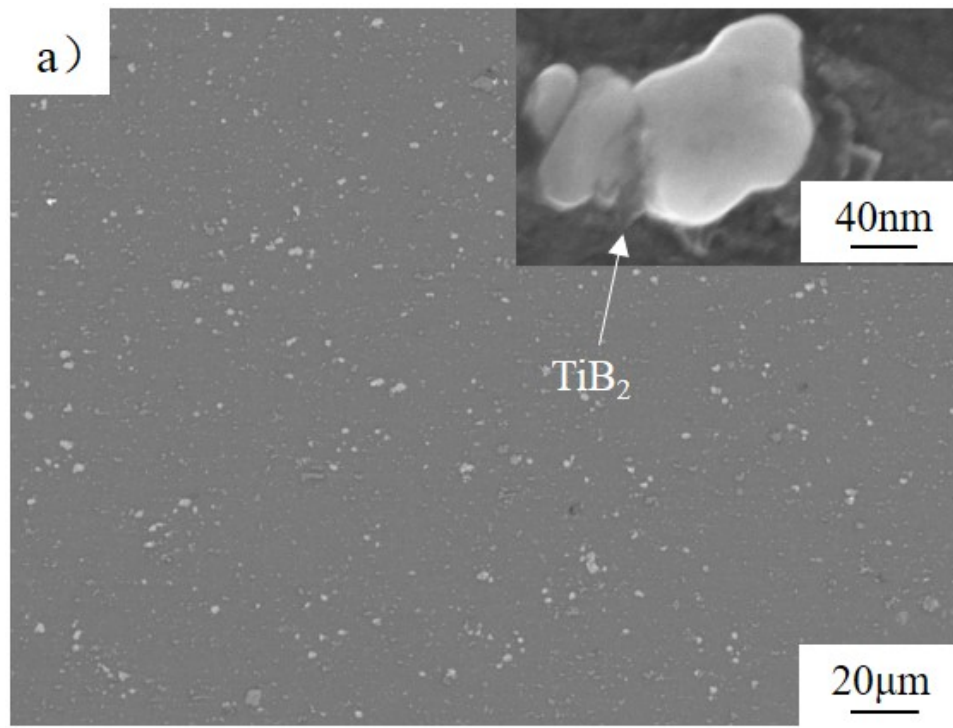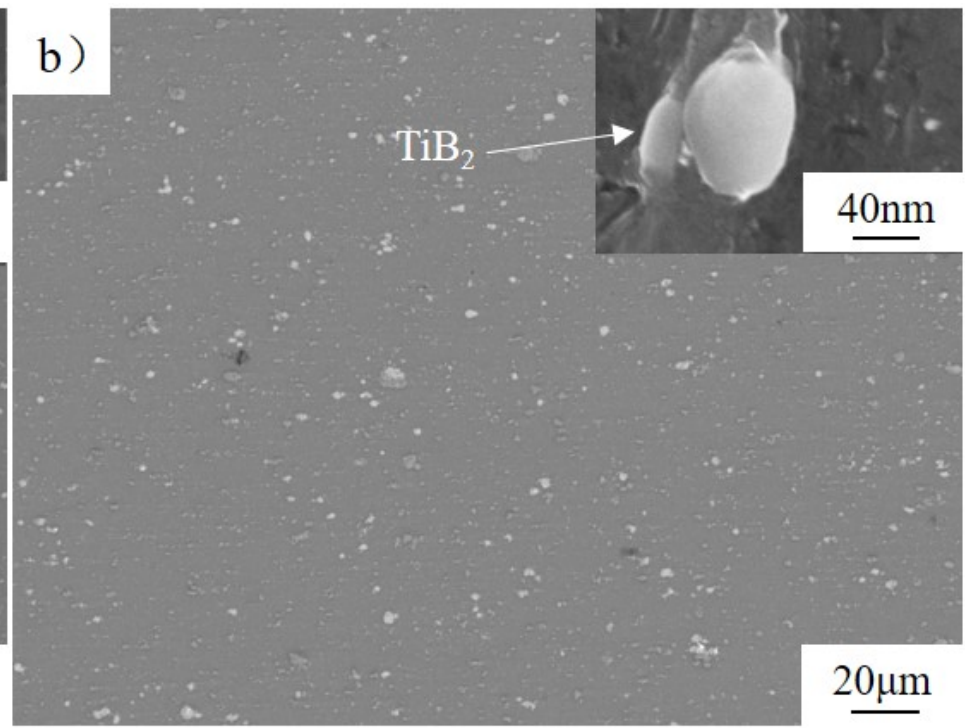

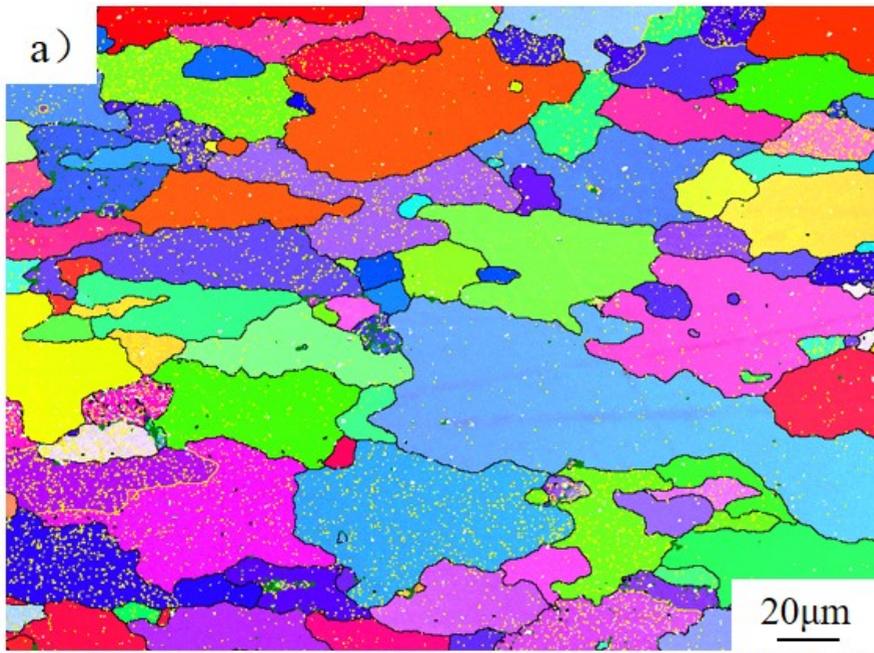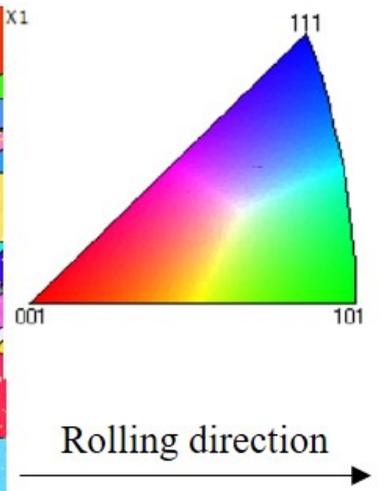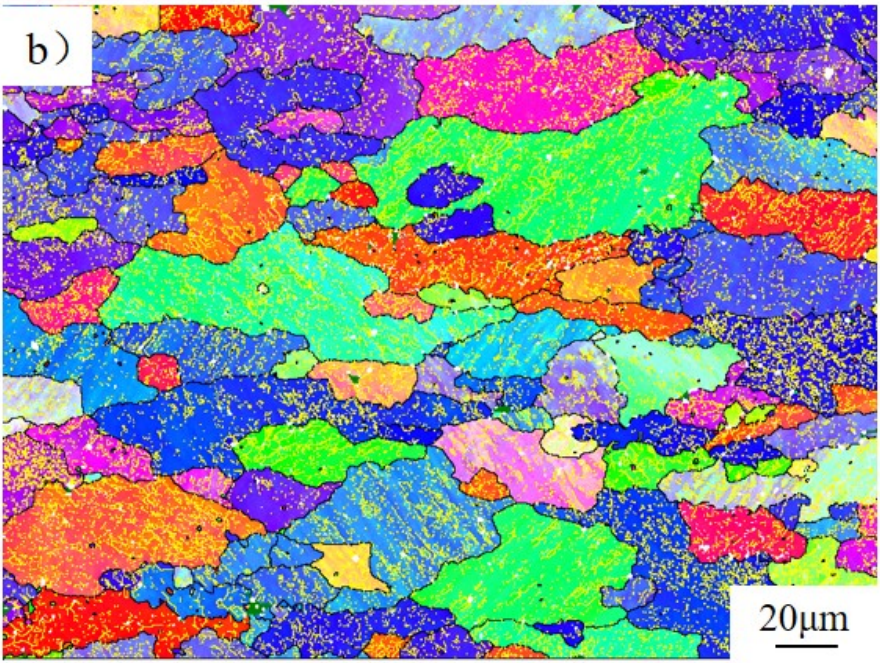

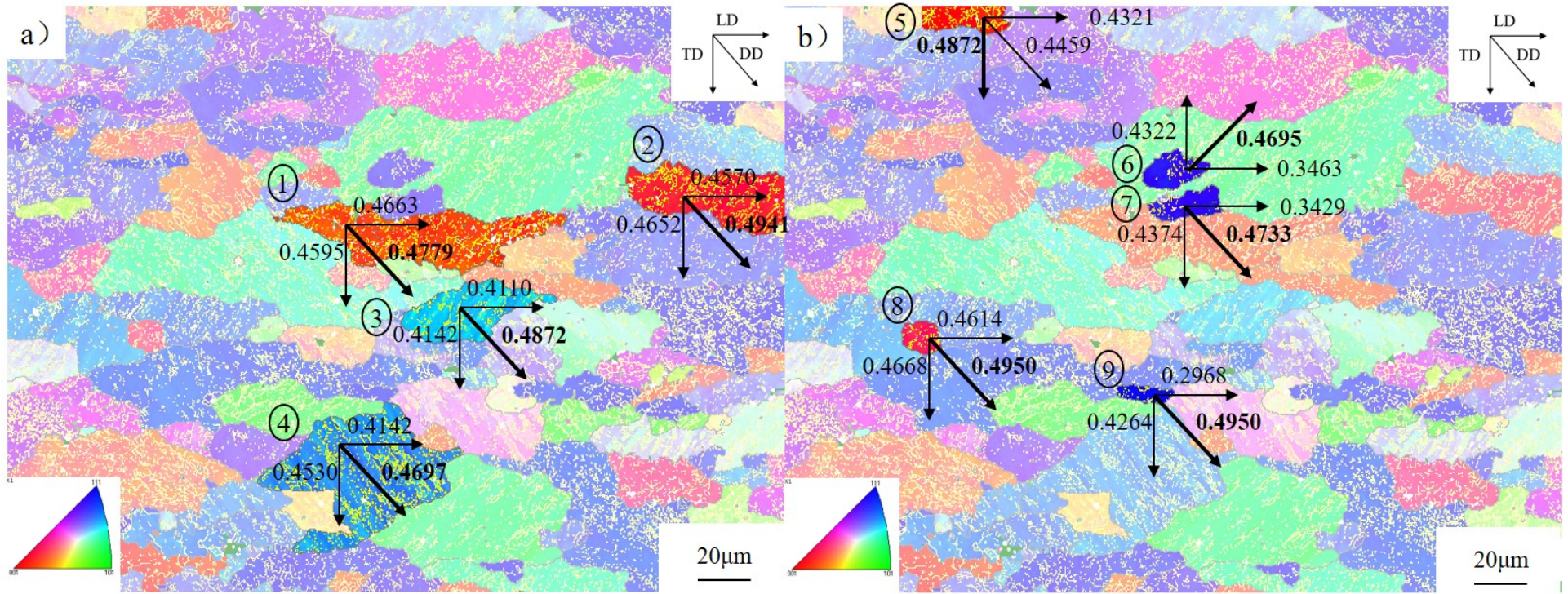

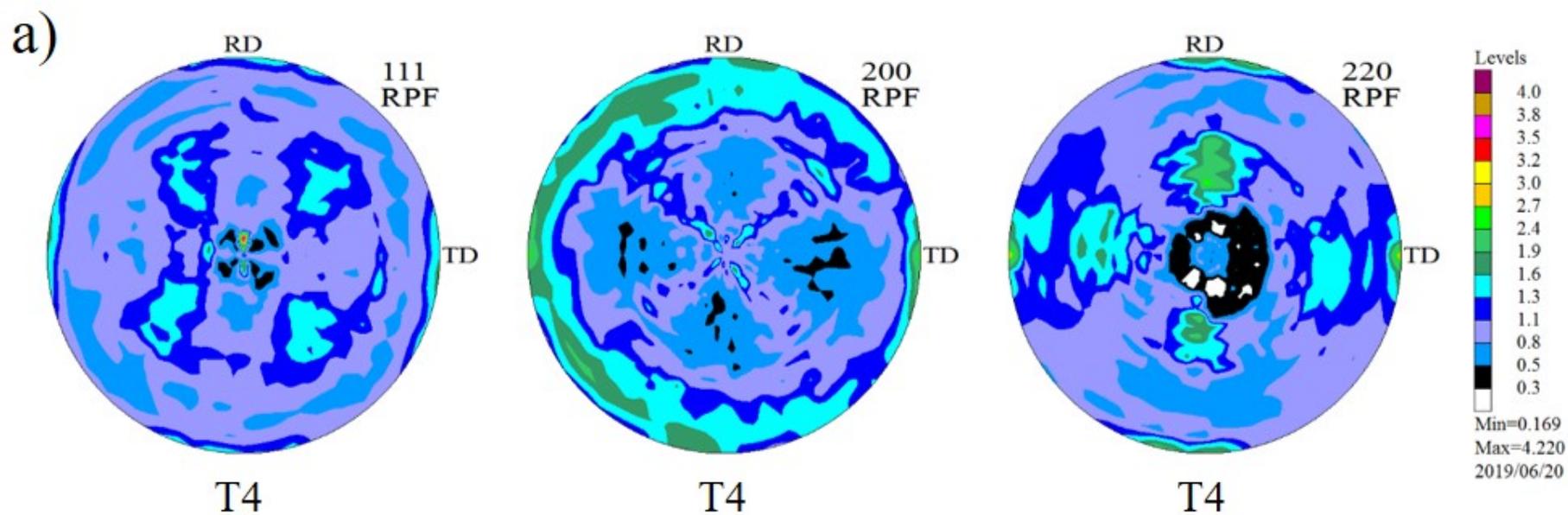
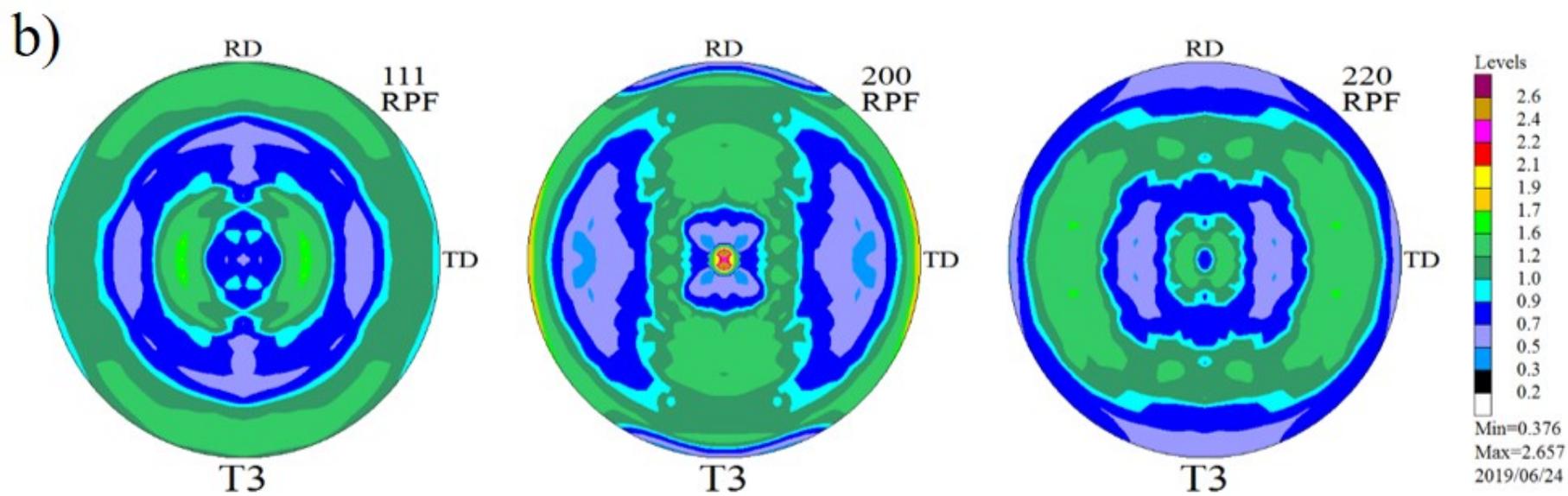

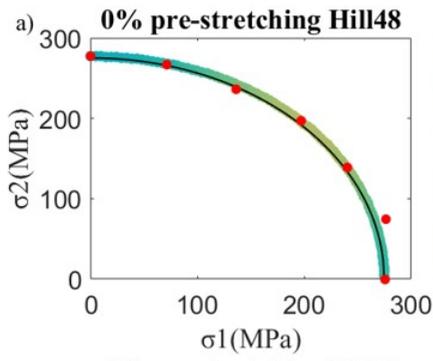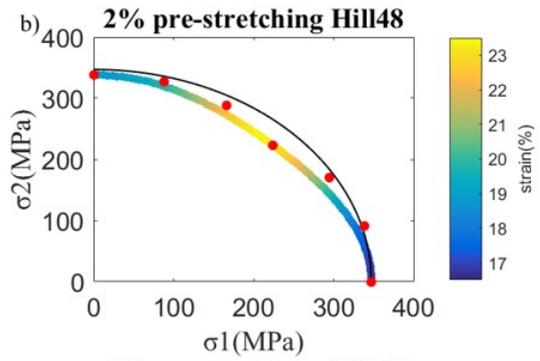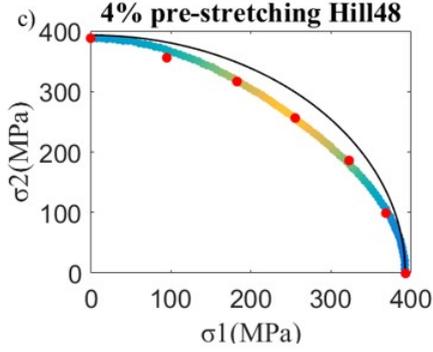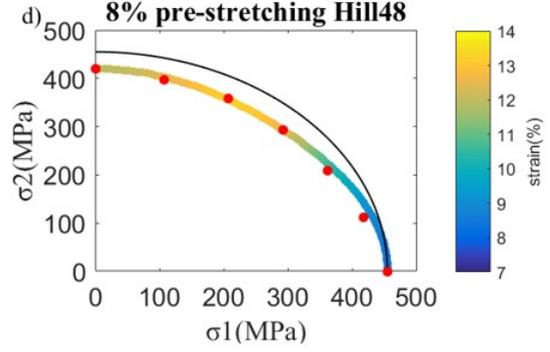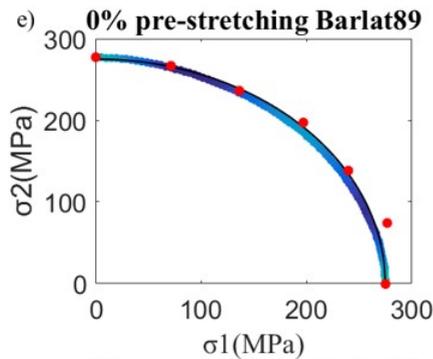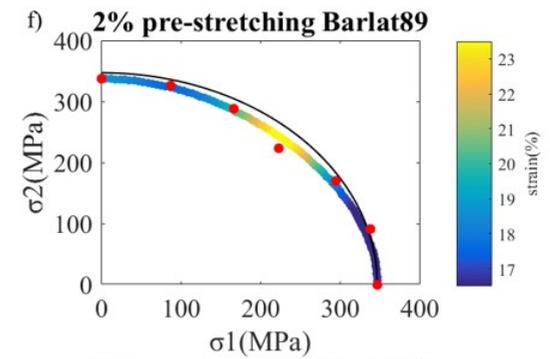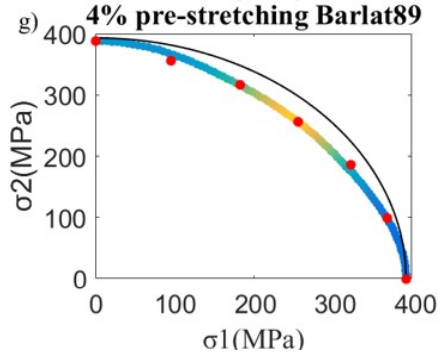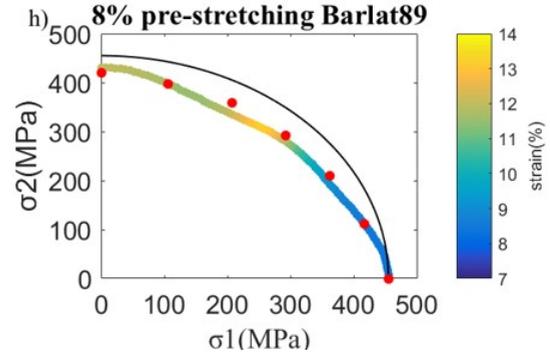

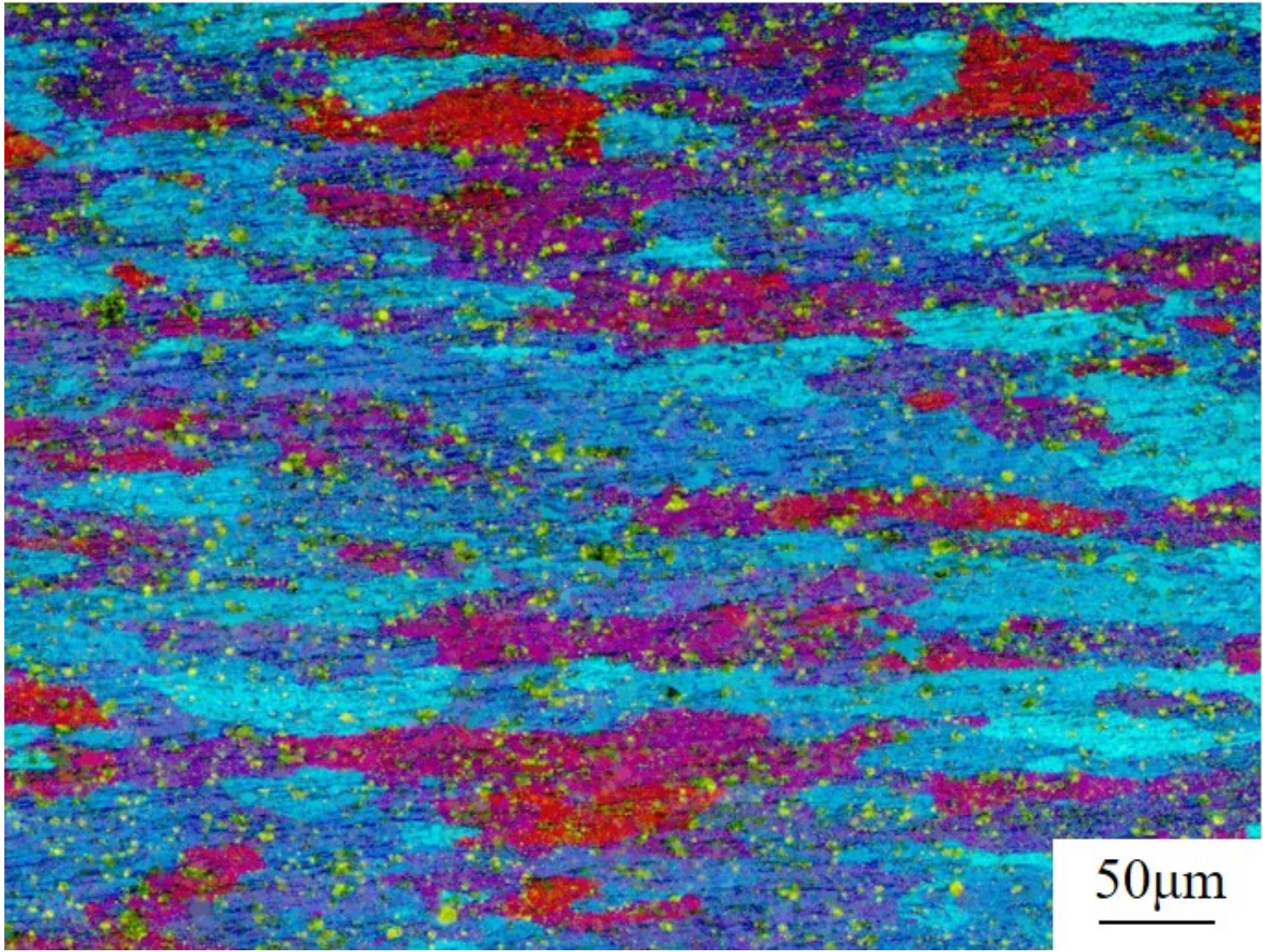

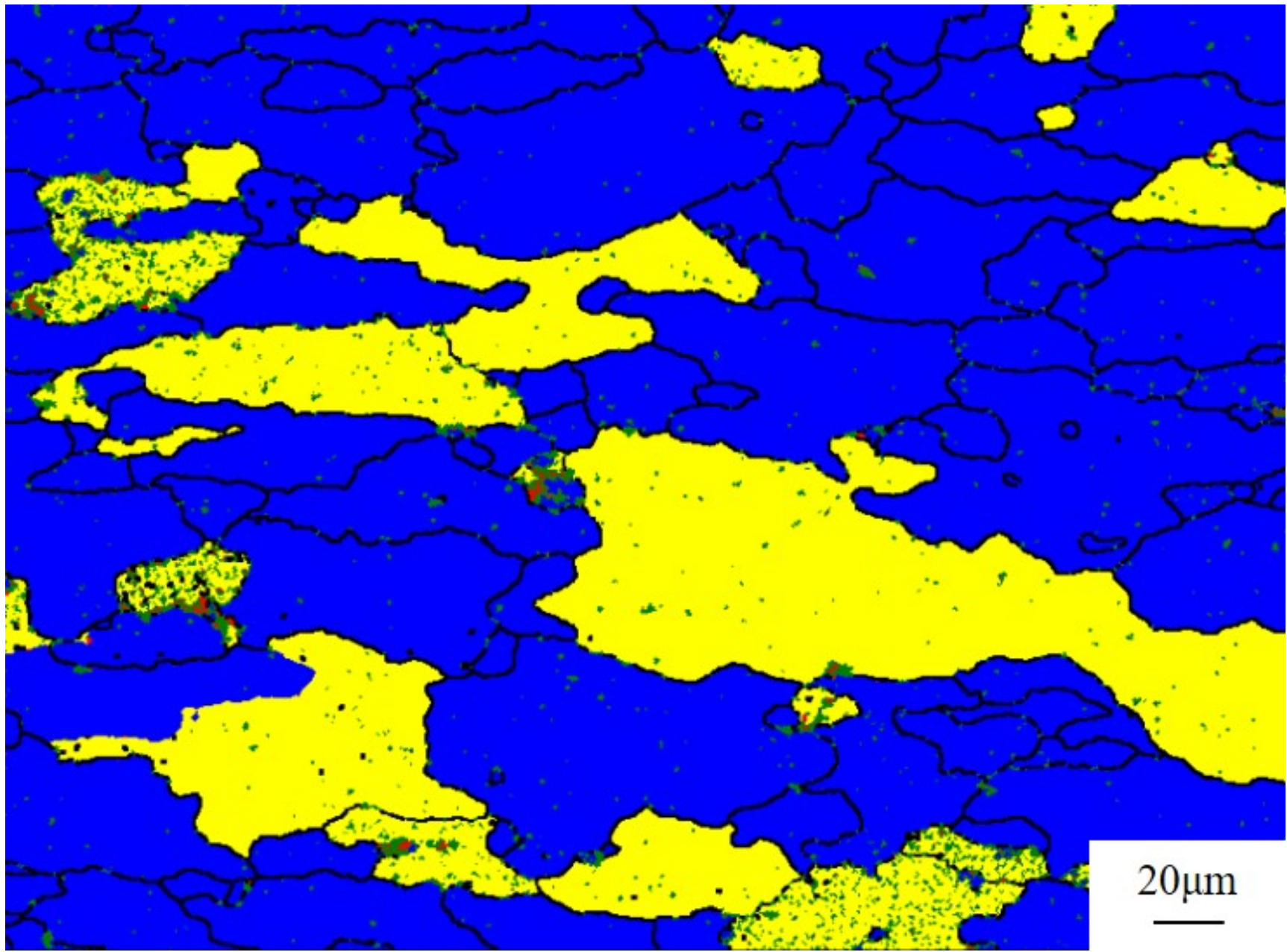

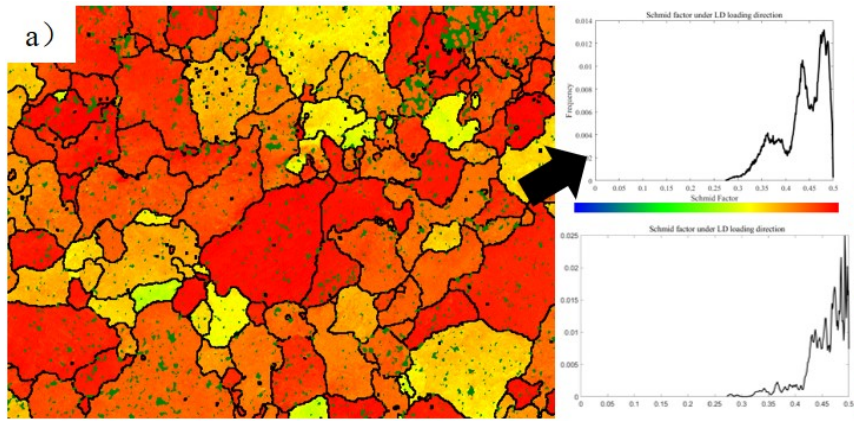
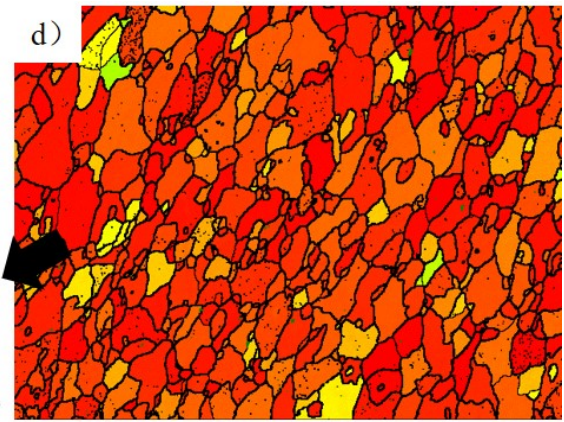
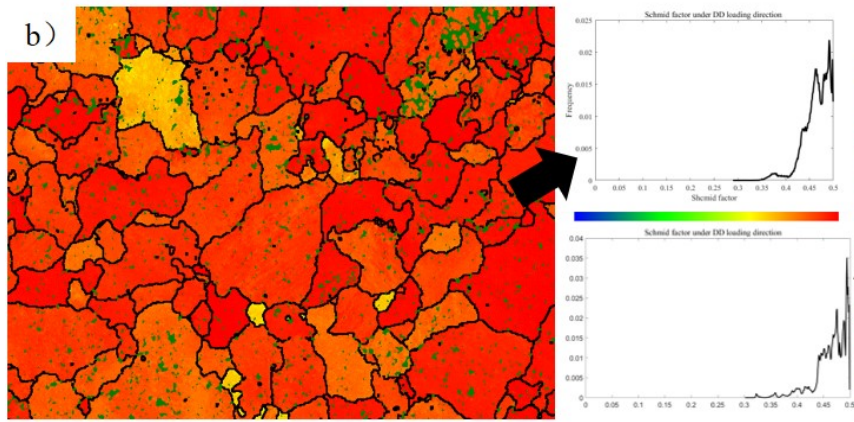
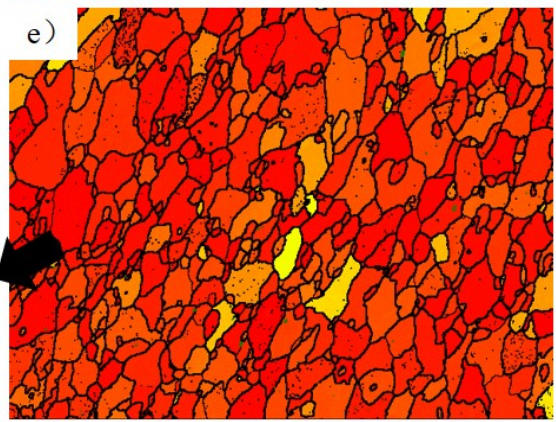
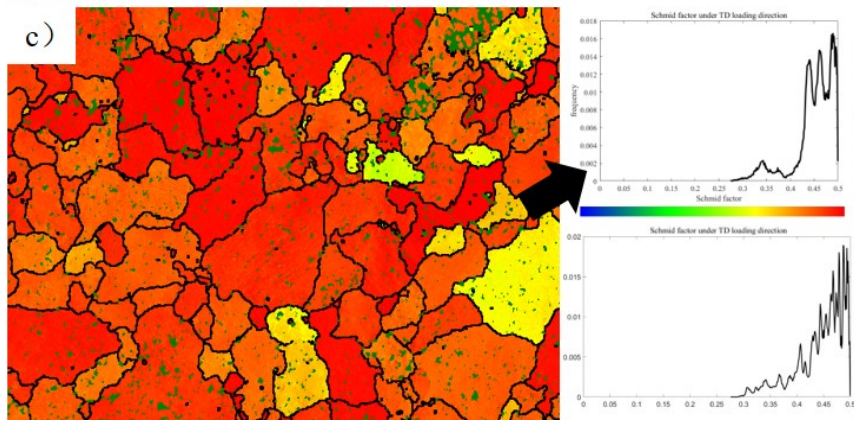
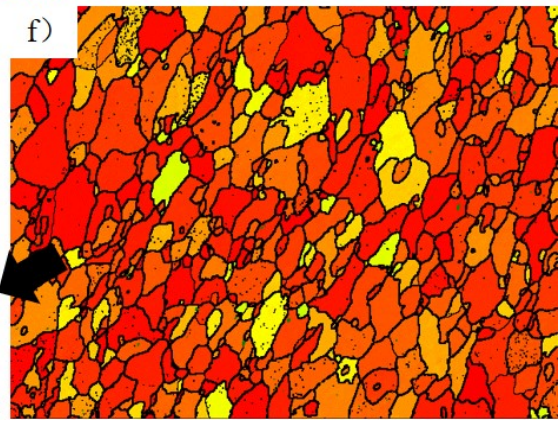

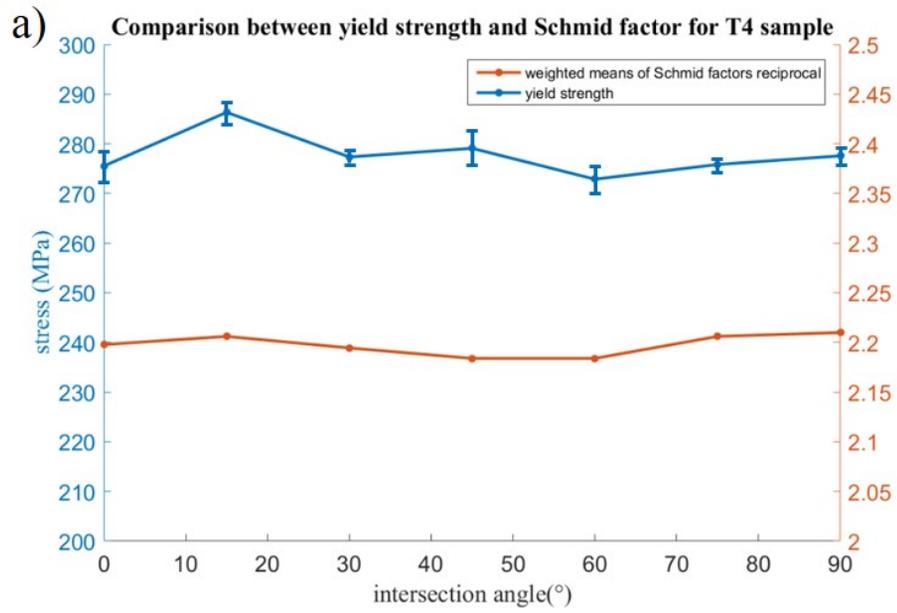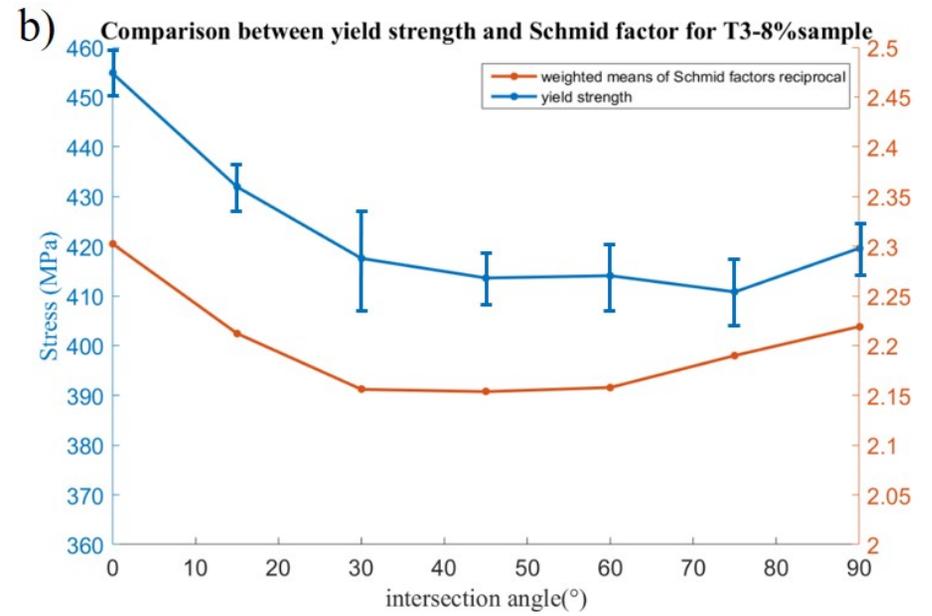

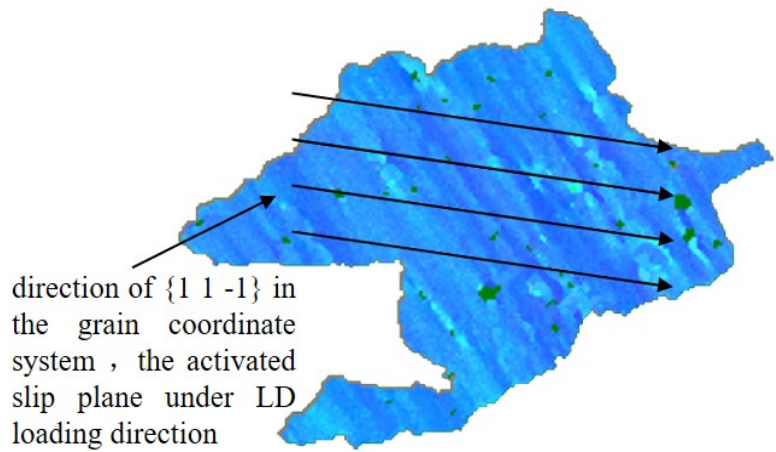
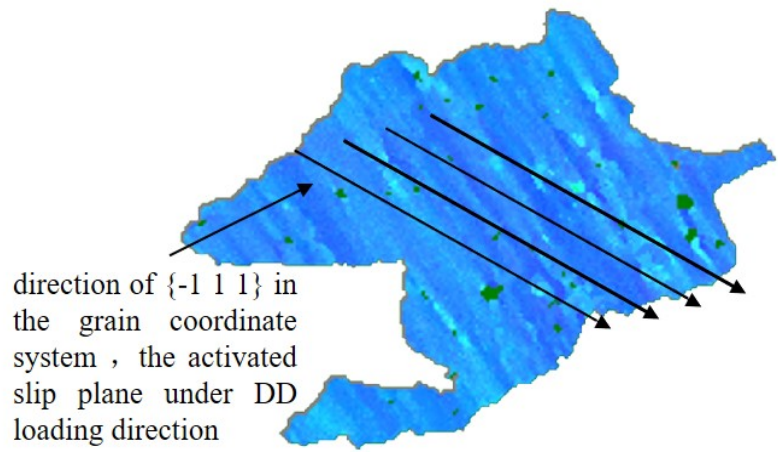
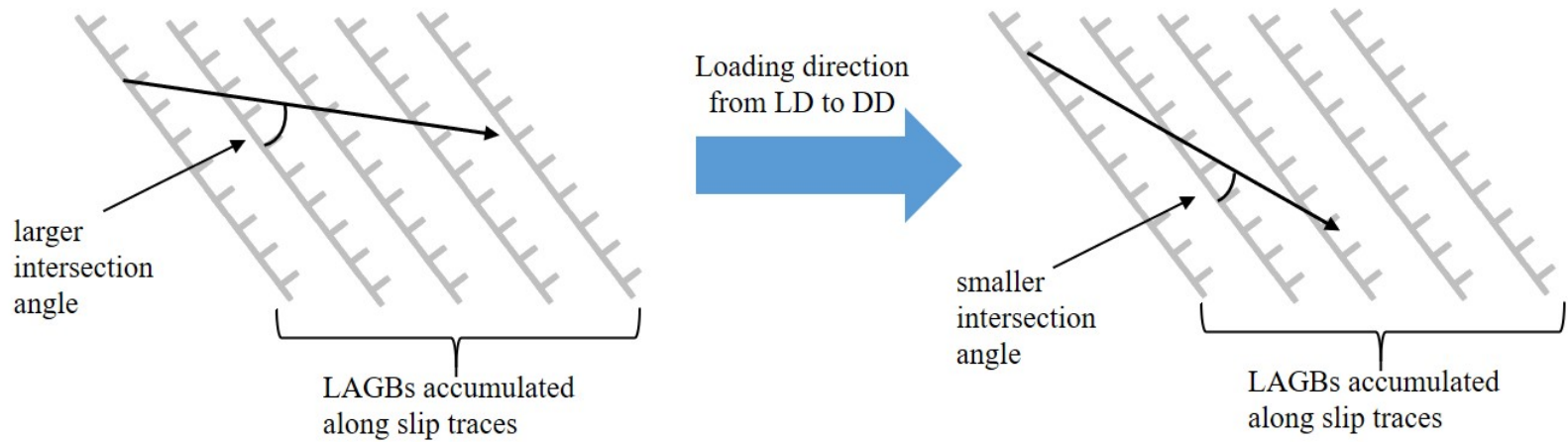